\documentclass[12pt,preprint]{aastex}

\shortauthors{Boboltz \& Marvel}
\shorttitle{OH\,12.8$-$0.9:  H$_2$O Maser Kinematics}

\begin{document}

\title{Water Maser Kinematics in the Jet of OH\,12.8$-$0.9}

\author{David A. Boboltz}
\affil{U.S. Naval Observatory, \\
3450 Massachusetts Ave., NW, Washington, DC 20392-5420, \\
dboboltz@usno.navy.mil}

\and

\author{Kevin B. Marvel}
\affil{American Astronomical Society, \\
2000 Florida Ave., NW, Suite 400, Washington, D.C., 20009-1231, \\
marvel@aas.org}

\begin{abstract}
We present Very Long Baseline Array observations of the kinematics 
of the water masers associated with OH\,12.8$-$0.9, the fourth 
member of the so-called ``water fountain" class of sources.  We find that 
the masers occupy two distinct regions at the ends of a bipolar jet-like structure 
oriented north--south, with the blue-shifted masers located to the 
north and the red-shifted masers to the south.  The masers are distributed 
along arc-like structures 12--20\,mas across oriented 
perpendicular to the separation axis with an angular separation of 
$\sim$110\,mas on the sky.  Our multi-epoch observations, show 
the two maser arcs to be expanding 
away from each other along the axis of separation.  The relative proper 
motions of the two maser regions is 2.7\,mas\,yr$^{-1}$ ($\sim$105\,km\,s$^{-1}$ at 
the assumed distance of 8~kpc).  The measured radial velocity difference
between the northern, blue-shifted masers and the southern, red-shifted 
masers is 48.4\,km\,s$^{-1}$.  The radial velocity, when combined with 
the proper motion, yields a three-dimensional expansion velocity of 
58\,km\,s$^{-1}$ and an inclination angle of 24$^{\circ}$ for the jet.  
By combining our radial velocities with historical values, we estimate the three dimensional acceleration of the masers to be $\sim$0.63\,km\,s$^{-1}$\,yr$^{-1}$ 
and a dynamical age for the collimated outflow of $\sim$90\,yr.  
   
\end{abstract}


\section{INTRODUCTION}

In the study of stars and stellar evolution, certain stages in the 
stellar life cycle remain a mystery simply because of their transitory 
nature.  The evolutionary stage between the end of the asymptotic
giant branch (AGB) and the planetary nebula (PN) phase is one 
such example.    As stars evolve up the AGB, they lose mass
at an increasing rate and spherically symmetric circumstellar envelopes 
(CSEs) are formed.  Despite the spherical symmetry of the mass-loss 
process, a large fraction ($\sim$75\%) of planetary nebulae (PNs) show 
aspherical (i.e. elliptical, bipolar, quadrupolar) morphologies 
\citep{MVSG:00}.  The shaping of asymmetrical PNs must therefore 
occur over the relatively short time period of 10$^3$--10$^4$\,yr that 
the star spends in a post-AGB or proto-planetary nebula (PPN) 
phase \citep{KWOK:93}.   Based on comparative studies of compact
and extended PNs, \cite{AK:96} find that the lack of difference in the 
morphologies between the two groups suggests that PNs morphologies
are primarily inherited from the AGB evolutionary stage.  \cite{ST:98} 
surveyed a number of young PNs with the Hubble Space Telescope 
Wide Field Planetary Camera 2 and found that the majority of them were
characterized by multi-polar bubbles distributed in a point-symmetric 
fashion about the central star.  They also found collimated radial structures 
and bright equatorial structures indicating the presence of jets 
and disks/torii in some objects.  These observations led \cite{ST:98}
to propose a mechanism in which high-speed collimated 
outflows operate during the late-AGB or early PPN stages of
stellar evolution.  During this phase, the jet-like outflows carve out
an imprint in the spherical AGB wind, and it is this imprint that 
provides the morphological signature necessary for the development of 
aspherical PNs.  
  
There is a small number of evolved objects that have been shown to 
exhibit the types of collimated jet-like outflows described above.  
These objects, dubbed ``water fountain" nebulae, are thought to be 
stars entering the post-AGB or PPN evolutionary stage.   
They exhibit both H$_2$O and OH maser emission, 
however, the relative characteristics of the two maser species differ from 
those of the typical AGB star.  These differences are apparent in both the 
spectral profiles of the two types of masers and in the unique spatial 
morphologies of the masers as measured by radio interferometry.
Prior to our study of OH\,12.8$-$0.9 \citep{BM:05}, there were three 
confirmed water-fountain sources: IRAS\,16342$-$3814 
\citep{STMZL:99, MSC:03}, IRAS\,19134+2131 \citep{IMAI:04}, and W43A 
\citep{IMAI:02}.  

Spectrally, these water-fountain sources exhibit double-peaked OH and 
H$_2$O maser profiles with the peaks symmetrically distributed about 
the radial velocity of the star.  However, unlike the typical evolved star, 
the H$_2$O maser peaks have a greater spread in velocity than the 
corresponding OH masers.   For  IRAS\,16342$-$3814, IRAS\,19134+2131, 
and W43A these velocity ranges for the H$_2$O masers are 
259\,km\,s$^{-1}$, 132\,km\,s$^{-1}$ and 180\,km\,s$^{-1}$ respectively 
\citep{LMM:92}.  In addition, the expansion velocities ($>$60\,km\,s$^{-1}$) 
implied by the maser spectra are much greater than the expansion velocity 
for the circumstellar wind of a typical AGB star (5--30\,km\,s$^{-1}$).
 
Spatially, the H$_2$O masers associated with these sources
are known to trace bipolar jet-like outflows, hence the "water fountain" name.  
In the first Very Long Baseline Interferometry 
(VLBI) study of the H$_2$O masers toward W43A, \cite{IMAI:02}
showed that the water masers are formed in a collimated precessing jet
with an estimated 3-D outflow velocity of 145\,km\,s$^{-1}$.   Interferometric 
studies of IRAS\,16342$-$3814 \citep{MSC:03, CSM:04} and IRAS\,19134+2131 
\citep{IMAI:04} also found the H$_2$O masers to exhibit bipolar distributions.  
Three-dimensional outflow velocities for IRAS\,16342$-$3814 and 
IRAS\,19134+2131 are  185\,km\,s$^{-1}$ and 130\,km\,s$^{-1}$, respectively.
The bipolar jets traced by the H$_2$O masers, presumably along the polar 
axis of the star, may represent the onset of the axisymmetric morphologies 
that typify PNs.  The dynamical ages of the jets for IRAS\,19134+2131 and 
W43A are estimated to be $\sim$50\,yr and $\sim$40\,yr, respectively 
\citep{IMAI:02, IMAI:04}.  That for IRAS\,16342$-$3814 is estimated to 
be $\sim$150\,yr \citep{CSM:04}.  

The enigmatic source OH\,12.8$-$0.9 was first classified by \cite{BAUD:79}
as a Type II OH/IR star based on its characteristic double-peaked 1612-MHz OH 
maser profile.   The distance to the source is unknown although 
\cite{BAUD:85} included it in a sample of OH/IR stars associated 
with the Galactic center.  OH\,12.8$-$0.9 has been linked to the infrared 
source IRAS\,18139$-$1816, which is $\sim$26\,arcsec away 
\citep{teLINTEL:89}.  The SIMBAD astronomical database, however, 
still lists two separate entries for the source, one for each identifier.  
Classification of IRAS\,18139$-$1816 based on IRAS Low Resolution 
Spectrometer (8--23\,$\mu$m) data was performed by \cite{KVB:97} who 
placed the source in category {\em I}, a group with noisy or incomplete 
spectra.  \cite{KVB:97} also noted that IRAS\,18139$-$1816 is a 
``25\,$\mu$m peaker", or a source whose IRAS flux density at 
25\,$\mu$m is greater than its flux density at both 12 and 60\,$\mu$m.  
Other 25\,$\mu$m peakers include young planetary nebulae and carbon 
stars with circumstellar silicate emission features \citep{KVB:97}.   
OH\,12.8$-$0.9 has been also   
observed with the PHT-S (2.5--11.6\,$\mu$m) spectrophotometer 
onboard the Infrared Space Observatory (ISO).  From the spectra, 
\cite{HKPW:04} classify the source in the {\em 4/5.SA} category.  
Sources in the {\em 4/5} group peak long-ward of the 11.6\,$\mu$m 
limit of the PHT-S.  Objects in the {\em 4/5.SA} sub-group additionally 
exhibit a deep 10\,$\mu$m absorption feature and absorption features from 
H$_2$O at 3 and 6\,$\mu$m, CO$_2$ at $\sim$4.3\,$\mu$m, and 
CO at $\sim$4.6\,$\mu$m \citep{HKPW:04}.  

H$_2$O maser emission from OH\,12.8$-$0.9 was first 
detected by \cite{ESW:86} who noted the fact that 
the H$_2$O emission peaks are outside the range of the OH maser emission.  
\cite{GOMEZ:94} used the Very Large Array (VLA) to show that the 
``anomalous" H$_2$O and OH maser emission was spatially coincident to within 
1$''$, and therefore belonged to the same source.  \citet{GOMEZ:94} also 
found double-peaked profiles for both the OH and the H$_2$O with the 
peaks of the OH separated by $\sim$23\,km\,s$^{-1}$ and H$_2$O peaks 
separated by nearly twice this amount, $\sim$42\,km\,s$^{-1}$.  
Although the velocity range for the H$_2$O is not as wide as the other 
water-fountain sources, \cite{ENGELS:02} found that the shape and 
variations in the spectra were consistent with the water-fountain class and 
compatible with an axisymmetric wind.  

In \cite{BM:05} we reported on our initial Very Long Baseline Array (VLBA)
observations of the H$_2$O masers associated with 
OH\,12.8$-$0.9.  We found that the H$_2$O masers trace the bipolar 
morphology typical of the water-fountain class.  This bipolar structure 
combined with the spectral characteristics of the H$_2$O and OH
maser emission led us to propose that OH\,12.8$-$0.9 is 
a fourth member of the rare water-fountain class of objects.    In this article, we 
present additional multi-epoch observations made with the VLBA with the 
goal of measuring the kinematics of the H$_2$O masers that trace 
the jet of OH\,12.8$-$0.9.

\section{OBSERVATIONS \label{OBS}}

We observed the 22.2 GHz H$_2$O maser emission from OH\,12.8$-$0.9 
($\alpha = 18^h 16^m 49^{s}.23, \delta = -18^{\circ} 15' 01''.8$, J2000)
using the 10 stations of the VLBA.  The VLBA
is operated by the National Radio Astronomy Observatory
(NRAO).\footnote{The National Radio Astronomy Observatory is a
facility of the National Science Foundation operated under cooperative
agreement by Associated Universities, Inc.}.   
VLBA spectral-line observations occurred on 2004 June 21, 2005 July 21,
2005 Nov. 2, and 2006 Feb. 24. 
Each observing run was approximately 5 hrs in length.  A reference frequency
of 22.23508 GHz was used for the H$_2$O maser transition.  
Data were recorded in dual circular polarization using 
two 8-MHz (112.6\,km\,s$^{-1}$) bands centered on the local standard 
of rest (LSR) velocity of $-$58.0\,km\,s$^{-1}$.  

For an unknown reason, the observations made on 2005 July 21 did 
not detect the H$_2$O masers from OH\,12.8$-$0.9.  Following this 
non-detection, we requested a short (5 minute) observation of OH\,12.8$-$0.9 
with the Green Bank Telescope (GBT) in order to decide whether the 
masers would still be detectable in the pending two epochs of VLBA 
observations.  On 2005 Sept. 13 the GBT observed the H$_2$O maser 
spectrum and found masers at levels of 2--3\,Jy.  We therefore continued 
with remaining two scheduled VLBA epochs and subsequently managed 
to detect and image the masers.  The remainder of this article relates 
only to the three epochs in which the maser observations were successful, 
hereafter denoted:  epoch 1 (2004 June 21), epoch 2 (2005 Nov. 2), 
and epoch 3 (2006 Feb. 24).

The data were correlated at the VLBA correlator operated by NRAO in 
Socorro, New Mexico.  Auto and cross-correlation
spectra consisting of 512 channels with channel spacings of 15.63\,kHz 
($\sim$0.22\,km\,s$^{-1}$) were produced by the correlator.  
Calibration of each of the three epochs was performed in accordance with 
standard VLBA spectral-line procedures using the Astronomical Image Processing 
System (AIPS) maintained by NRAO.  The calibration of epoch 1 
is described in \cite{BM:05} and the two subsequent epochs were calibrated
in a similar manner.  For each epoch, residual delays due to the instrumentation 
were corrected by performing a fringe fit on the continuum calibrator (J1751+0939)
scans.   Residual group delays for each antenna were determined and applied to the 
target data.  

The bandpass response was determined from scans on J1751+0939
and was used to correct the OH\,12.8$-$0.9 data.
The time-dependent gains of all antennas relative to a reference
antenna were determined by fitting a total-power spectrum (from the
reference antenna with the target source at a high elevation) to the
total power spectrum of each antenna. The absolute flux density
scale was established by scaling these gains by the system temperature and
gain of the reference antenna.  Errors in the gain and pointing of
the reference antenna and atmospheric opacity variations contribute to
the error in the absolute amplitude calibration, 
which is accurate to about 15--20\%.

Residual fringe rates were obtained by fringe-fitting a strong reference 
feature in the spectrum of OH\,12.8$-$0.9.  For each epoch we 
used the same strong feature at channel velocity 
$V_{\rm LSR} = -81.6$\,km\,s$^{-1}$.  
The resulting fringe-rate solutions were applied to all channels in the 
spectrum.  An iterative self-calibration and imaging procedure was then 
performed to map the emission in the reference channel for each epoch.  
The resulting residual phase and amplitude corrections from the reference 
channel were then applied to all channels in the 8-MHz band.  All of
the above calibrations were then applied to the data prior to imaging.  

Imaging of the epoch 1 data is discussed in \cite{BM:05}, and a similar
methodology was employed for the two subsequent epochs.
For epochs 2 and 3, full resolution images $2048\times 2048$ pixels 
($\sim$160$\times 160$\,mas) were generated using a synthesized beam
sizes of $1.08\times 0.39$\,mas for epoch 2 and $0.88\times 0.33$\,mas for
epoch 3.  Images were produced for all spectral channels 
from $-$88.3\,km\,s$^{-1}$ to $-$25.1\,km\,s$^{-1}$ forming 
an image cube of 301 image planes.  Since peak maser flux densities were 
on the order of 1 Jy, channel images are noise limited and do not suffer from 
dynamic range limitations as is often the case in maser observations.  
Typical 1$\sigma$ off-source noise estimates for individual image planes were 
8--10~mJy~beam$^{-1}$ for epoch 2 and 10--12~mJy~beam$^{-1}$ for 
epoch 3.  The analysis and extraction of relevant information from the image
cubes is described below.

\section{RESULTS}

In order to identify and extract maser component parameters, two-dimensional
Gaussian functions were fit to the emission in the image plane of each 
spectral (velocity) channel using the AIPS task SAD.  
Image quality was assessed using the off-source rms noise in the image.  
A cutoff flux density was set at five times the rms noise in the image plane
containing the emission.  Emission features with flux densities greater than 
the cutoff were fit with Gaussians to determine maser component parameters.  
The errors in right ascension and declination of the Gaussian fits to the identified 
components were computed by the AIPS task SAD following the method 
outlined by \citet{CONDON:97}.  These position errors ranged from 
5~$\mu$as for features with high signal-to-noise, to 110~$\mu$as for 
features with lower signal-to-noise.  
 
\begin{figure}[hbt]
\plotone{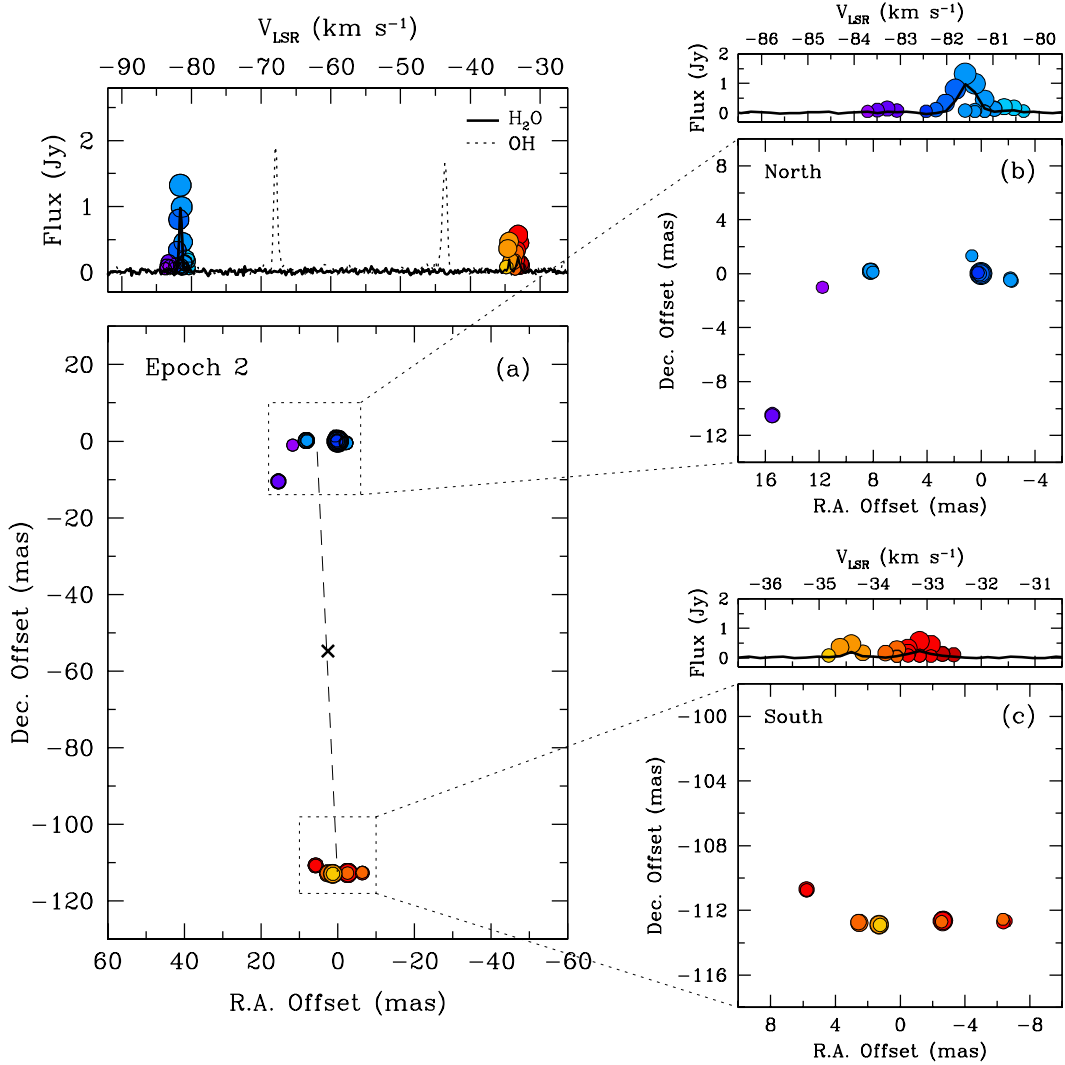}
\figcaption{\small The H$_2$O maser emission toward OH\,12.8--0.9 
from epoch 2 (2005 Nov. 2).  In panels (a), (b) and (c), the top sub-panels 
show the spectra formed by plotting maser component flux density versus 
LSR velocity, color-coded according to maser velocity.    A dark solid line 
in each upper sub-panel represents the scalar-averaged cross-power 
spectrum on the Los Alamos--Pie Town VLBA baseline.   A dotted line in 
the top sub-panel of (a) represents the scalar-averaged cross-power spectrum of 
the OH masers (with the flux density scaled by a factor of 0.5) from the 
Los Alamos--Pie Town baseline.  The lower sub-panels in (a), 
(b) and (c) plot the spatial distribution of the H$_2$O masers with point-color 
representing the corresponding velocity bin in the spectrum and 
point-size proportional to the logarithm of the maser flux density.   
Panel (a) shows all H$_2$O maser components from our VLBA observations.  
A dashed line represents the axis of separation between the centers of the 
blue- and red-shifted masers at a position angle of 2.6$^{\circ}$ east of north.  
The ``$\times$'' symbol represents the midpoint of the bipolar distribution.  
Panels (b) and (c) show expanded views of the blue- and red-shifted maser 
features respectively.  Errors in the positions of the features are smaller 
than the data points for all panels. \label{SPOT_MAP2}}
\end{figure}

\begin{figure}[hbt]
\plotone{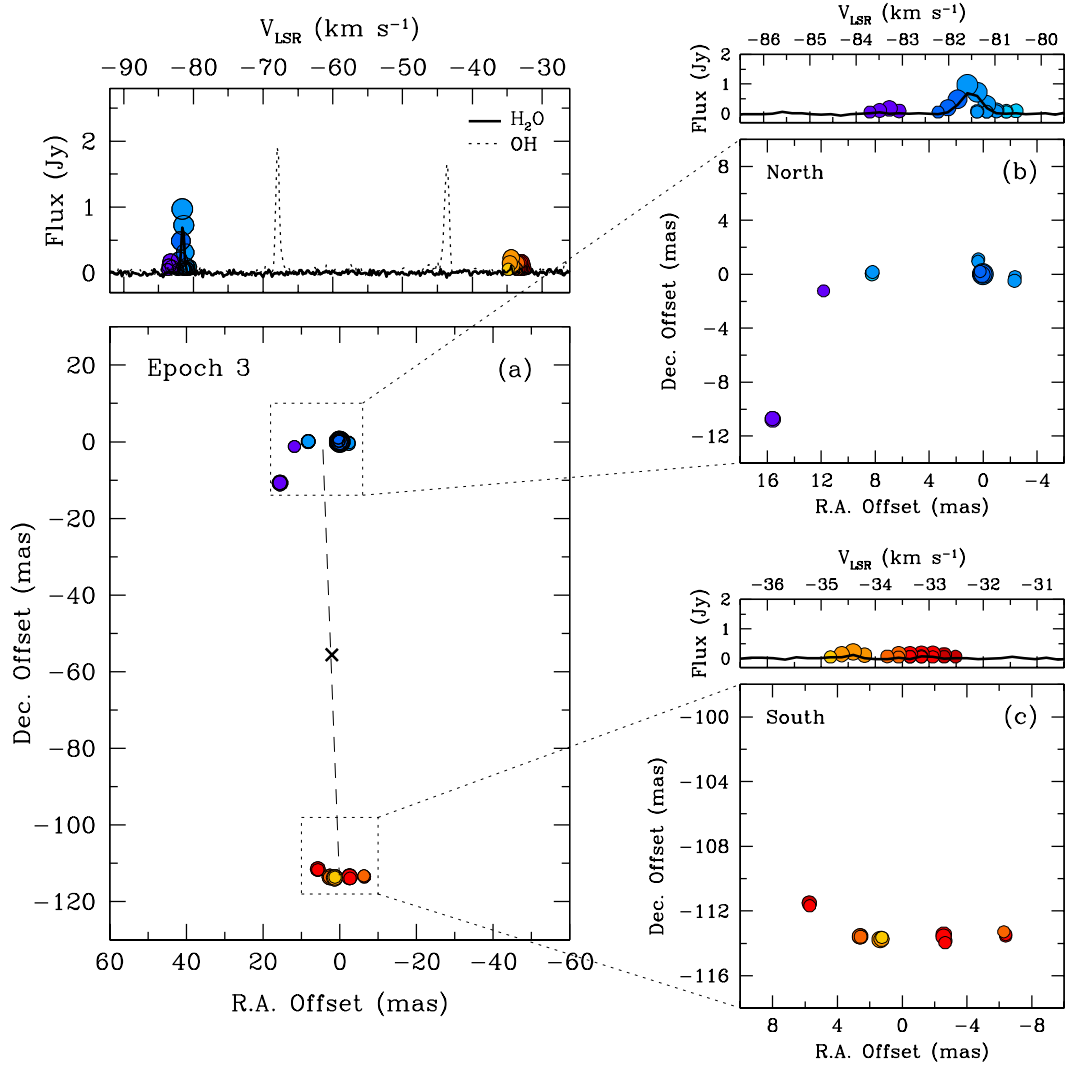}
\figcaption{Same as Figure \ref{SPOT_MAP2}, but for epoch 3 
(2006 Feb. 24). \label{SPOT_MAP3}}
\end{figure}

Figures~\ref{SPOT_MAP2} and \ref{SPOT_MAP3} show the spectral 
(upper sub-panels) and spatial (lower sub-panels) distributions of the 
H$_2$O masers toward OH\,12.8$-$0.9 from the analysis of our VLBA 
images for epoch 2 (Fig.~\ref{SPOT_MAP2}) and epoch 3 (Fig.~\ref{SPOT_MAP3}).
A similar plot for epoch 1 was presented in \cite{BM:05}.
In the figures, panel (a) shows the entire range of H$_2$O maser emission from 
the star.  Panels (b) and (c) show enlarged views of the blue-shifted masers to 
the north and the red-shifted masers to the south, respectively.   

In the upper sub-panels in Figures~\ref{SPOT_MAP2}(a) and 
\ref{SPOT_MAP3}(a), the H$_2$O maser spectra for epochs 2 and 3 
are plotted along with the spectrum of the 1612-MHz OH masers 
from previous observations \citep{BM:05} as a reference.   
The velocities of the OH maser peaks, have not changed significantly 
over time when compared to previous observations of \cite{GOMEZ:94}.
We observed the velocities of the blue- and red-shifted 
OH peaks to be $-$68.0 and $-$43.7\,km\,s$^{-1}$, respectively with 
a velocity resolution of 0.18\,km\,s$^{-1}$.  By comparison, the 
velocities of the two peaks reported in \cite{GOMEZ:94} are 
essentially the same at $-$67.4 and $-$44.0\,km\,s$^{-1}$, 
to within the velocity resolution of their VLA observations 
(1.1\,km\,s$^{-1}$).   As in epoch 1, the H$_2$O masers
form a double-peaked profile with a velocity extent greater than that of 
the double-peaked OH maser profile.  On the blue-shifted side of the 
emission, the peak flux density of epoch 2 remained the same as for 
epoch 1 at 1.3~Jy~beam$^{-1}$ and decreased slightly in epoch 3 to 
0.97~Jy~beam$^{-1}$.  For
the red-shifted emission, the peak flux density steadily decreased from 
1.0~Jy~beam$^{-1}$ in epoch 1 to 0.6~Jy~beam$^{-1}$ in epoch 2 
to 0.2~Jy~beam$^{-1}$ in epoch 3.  The velocity
distributions for the masers in epochs 2 and 3 are very similar to those
in epoch 1 and will be discussed in detail in \S~\ref{RAD_VEL}.

The lower sub-panels of Figures~\ref{SPOT_MAP2}(a) and 
\ref{SPOT_MAP3}(a) show the entire spatial extent of the H$_2$O
masers for epochs 2 and 3, respectively.  Enlarged views of the north
and south regions are also shown in panels (b) and (c) of each 
figure.  We find that in epochs 2 and 3 the overall spatial structure of the
H$_2$O masers has remained essentially unchanged from epoch 1.  
There are two distinct regions oriented roughly north--south on the 
sky separated by $\sim$110\,mas.  The position angle of the axis of 
separation between the mean position centers of the 
northern and southern maser regions is $\sim$2.6$^{\circ}$ east of north.  
The masers in the two regions are arranged in arc-like structures oriented 
roughly east--west with the exception of the feature located near 
(15.0, $-$10.0) in Figures~\ref{SPOT_MAP2}(b) and \ref{SPOT_MAP3}(b).  
This feature, although apparent in the epoch 1 image cube, did not 
meet all of the component selection criteria in our initial analysis of the 
data \citep{BM:05}.  The feature is discussed further in \S~\ref{PROP_MOT}
below.  The blue- and red-shifted arcs of masers are approximately 
20 and 12~mas across, respectively, corresponding opening angles for 
the jet of 13--20$^{\circ}$.  

\subsection{Maser Proper Motions \label{PROP_MOT}}

In Figures~\ref{SPOT_MAP2} and \ref{SPOT_MAP3}, we plotted the maser 
emission identified in every spectral channel, thus an individual maser feature 
will appear in multiple spectral channels and will consist of multiple points in 
the spatial distributions shown in the figures.  In order to track and estimate the 
motions of the maser features, it is necessary to determine average values 
in right ascension, declination, and velocity for each feature.  These average
values were computed using a flux-density-squared weighted average for
emission identified in two or more channels and spatially coincident 
to within 1~mas.  The flux assigned to the maser feature was simply the 
peak flux in the channels spanned by the feature.  Characteristics of the 
masers for the three epochs are presented in 
Tables~\ref{COMP_TABLE}--\ref{COMP_TABLE3}.
  
Since we did not use the technique of phase-referencing for our three
VLBA epochs, the absolute position of the phase center in each image 
cube is unknown.  However, in the reduction and analysis of each epoch 
of observations, we used what we believe to be the same maser feature 
as a phase reference.  This feature was the strongest feature in each of
the three epochs.  In addition, it maintained the same velocity and spatial 
location relative to the other three northern, blue-shifted masers common 
to all three epochs.  The averaging of the maser identifications over several 
channels resulted in slight ($<$0.1~mas) position shifts for the reference feature in  
each of the three epochs.   The coordinate frames for all three epochs 
were therefore realigned such that the reference feature (the third blue-shifted
feature in Tables~\ref{COMP_TABLE} and the second blue-shifted 
feature in Tables~\ref{COMP_TABLE2} and \ref{COMP_TABLE3})
coincided with the origin.   

\begin{figure}[hbt]
\plotone{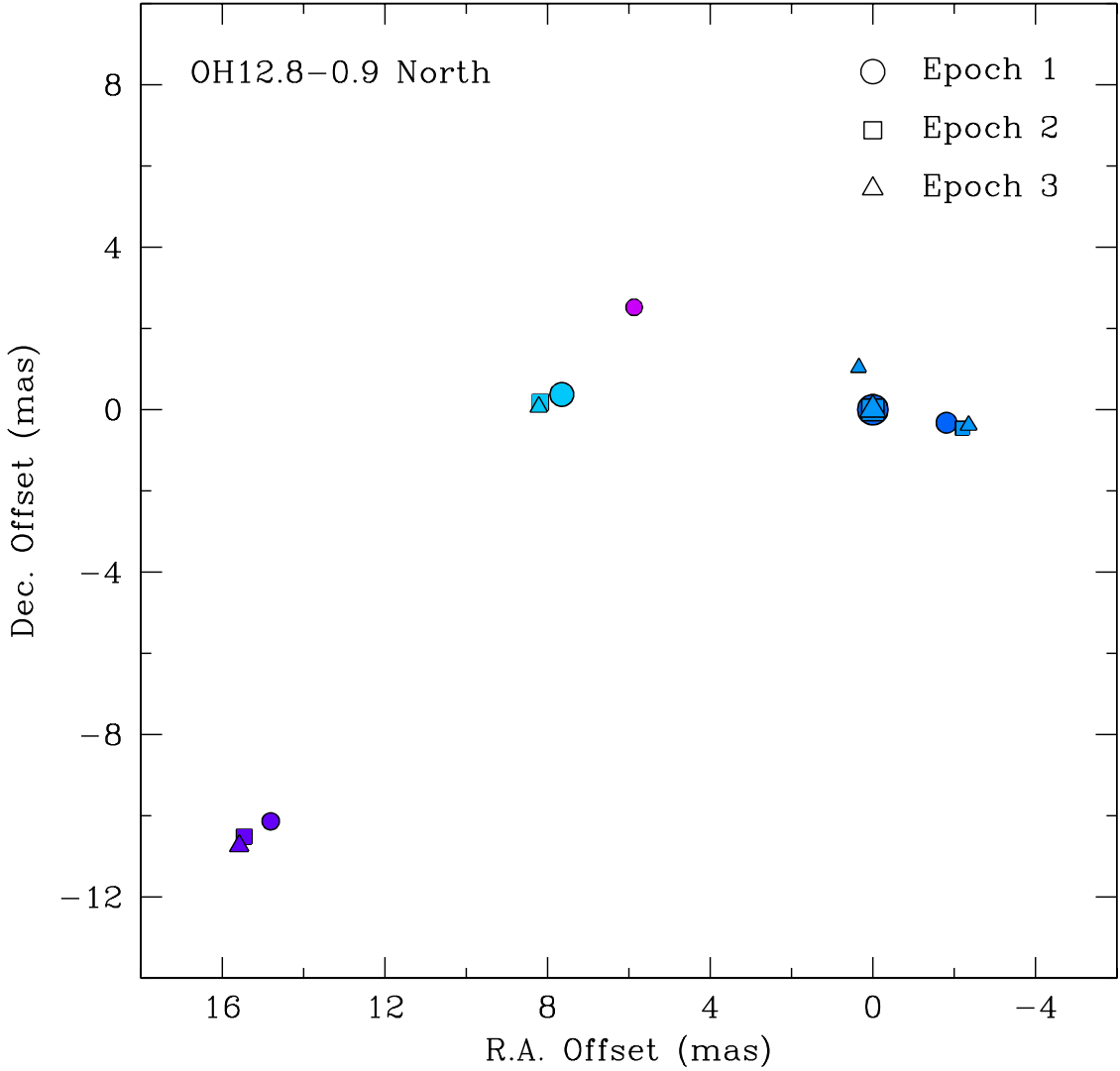}
\figcaption{Northern blue-shifted masers associated with OH\,12.8$-$0.9 
for all three epochs (2004 June 21, 2005 Nov. 2, and 2006 Feb. 24). 
Masers are color-coded according to velocity and point-size  
is proportional to the logarithm of the maser flux density.  There are four 
maser features common to all three epochs including the reference 
feature at (0,0). The masers show little motion relative to this stationary
reference feature.
\label{NORTH_COMPS}}
\end{figure}

\begin{figure}[hbt]
\plotone{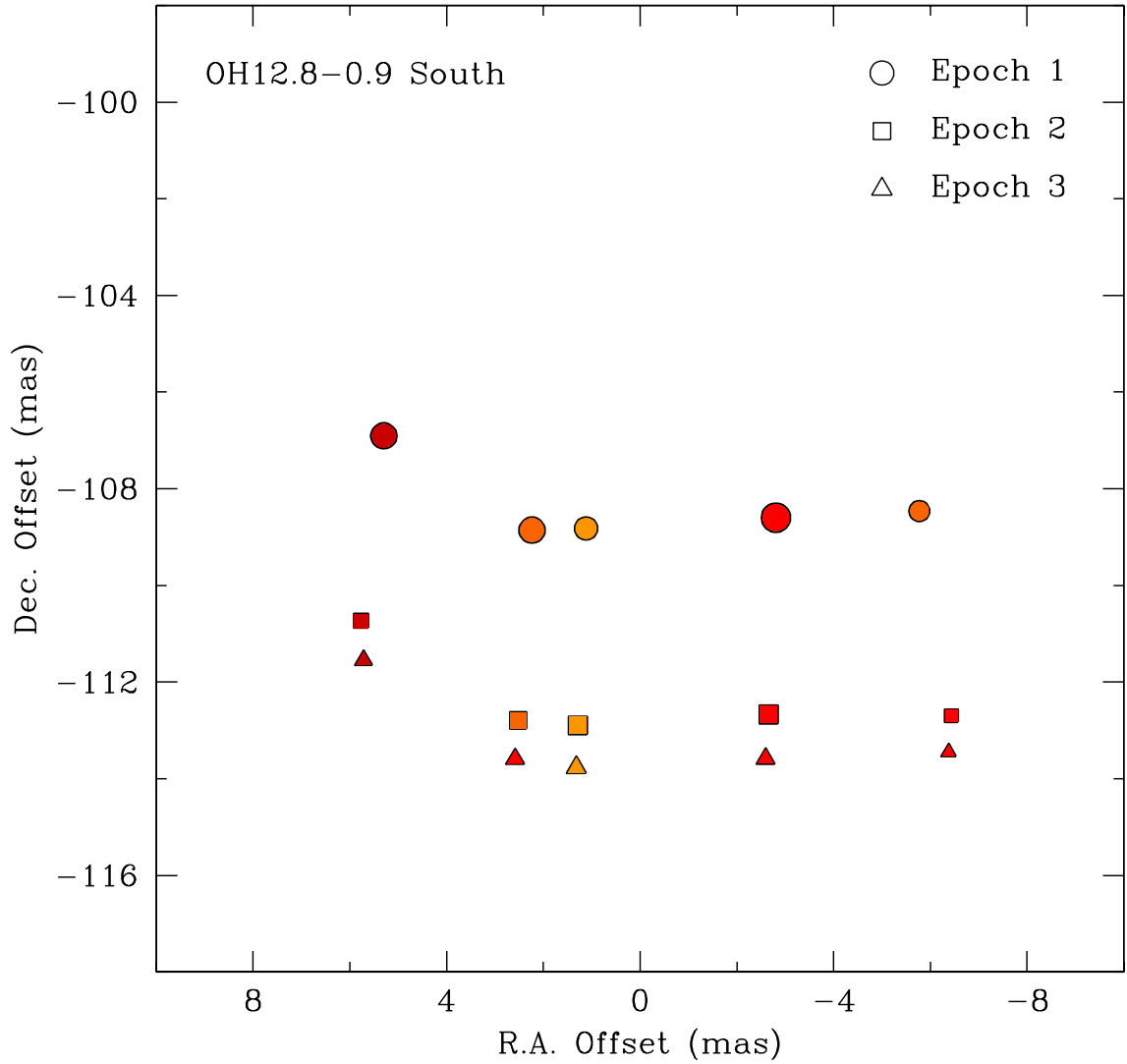}
\figcaption{Southern red-shifted masers associated with OH\,12.8$-$0.9 
for all three epochs (2004 June 21, 2005 Nov. 2, and 2006 Feb. 24)
demonstrating the motions of the masers over time relative to the 
reference feature in Figure \ref{NORTH_COMPS}.  Masers are again 
color-coded according to velocity and point-size is proportional
to the logarithm of the maser flux density.  All five features are common 
to all three epochs.
\label{SOUTH_COMPS}}
\end{figure}

Relevant maser characteristics determined from the analysis are plotted 
in Figures~\ref{NORTH_COMPS} and \ref{SOUTH_COMPS}.
Figure~\ref{NORTH_COMPS} shows the blue-shifted H$_2$O masers in the north
of OH\,12.8$-$0.9.  In this figure, the alignment of the reference features for 
the three epochs at the origin is apparent.  Also common to all three epochs 
are features near ($-$2.0, 0.5), (8.0, 0.2) and (15.0, $-$10.0).   
As mentioned earlier, the feature near (15.0, $-$10.0) did not appear in 
\cite{BM:05} because it did not meet all of the selection criteria at the 
time.  However, with the additional knowledge from epochs 2 and 3, 
and some reprocessing of the epoch 1 image cube, we were able to extract the 
relevant parameters for this feature.  From Figure~\ref{NORTH_COMPS} it is 
apparent that any position differences between the three epochs are small, 
thus indicating little motion of the northern features relative to the reference 
feature over the course of the three experiments.  This, however, is not the 
case for the southern maser features.  Figure~\ref{SOUTH_COMPS} shows 
the red-shifted H$_2$O masers in the south of OH\,12.8$-$0.9.  Here there are 
five features that are common to all three epochs.   Clearly there is a consistent
proper motion of all of the features nearly due south relative to the blue-shifted 
reference feature.  The average angular separations of the north and south 
masers are 107, 110 and 111~mas for epochs 1, 2 and 3, respectively. 
As discussed in \cite{BM:05} the distance to OH\,12.8$-$0.9 is not well known.  
Best estimates place the object near the Galactic center at $D \approx 8$\,kpc 
\citep{BAUD:85}.  Assuming this distance, then the linear separations between 
the blue- and red-shifted masers are approximately 860, 880 and 890~AU, 
respectively.   

\begin{figure}[hbt]
\plotone{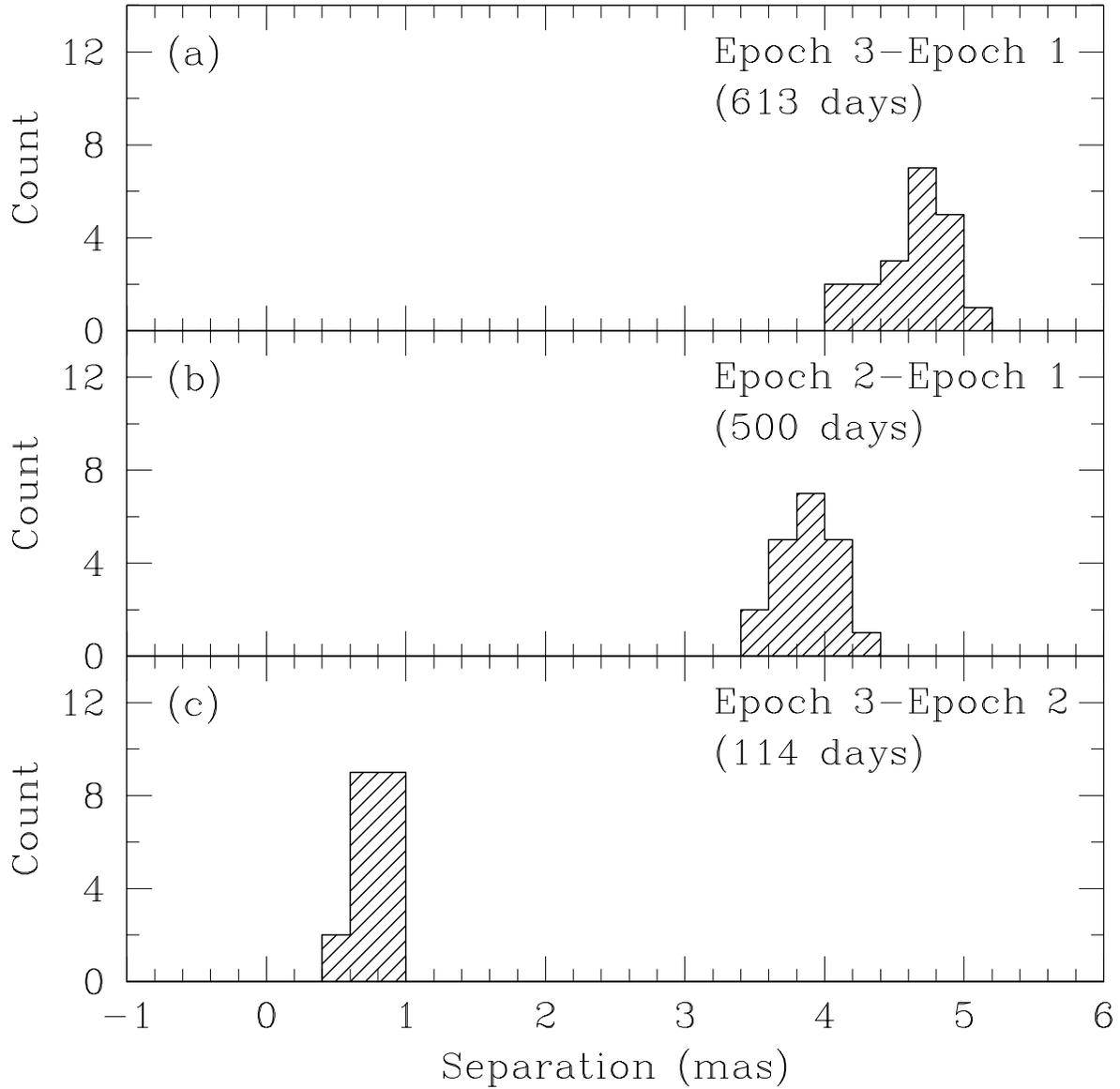}
\figcaption{Histogram showing the pairwise spatial separations between 
combinations of northern and southern maser pairs for (a) epochs 3 and 1, 
(b) epochs 2 and 1, and c) epochs 3 and 2. Listed in each panel is the 
elapsed time between epochs in days.  \label{PROP_HIST}}
\end{figure}

In order to better characterize the net expansion of the masers we computed 
separations between pairwise combinations of components.  This technique has 
previously been used to characterize OH \citep{CCS:91, BRM:92, KEMBALL:92}, 
H$_2$O  \citep{MARVEL:96}, and SiO \citep{BDK:97, CSIK:06} maser motions 
and has no dependence on the alignment of the maps or a priori assumptions
about the velocity field.  The procedure involves computing the 
angular separation ($\Delta \theta$) between two features at one epoch 
(epoch~A) and the separation between the corresponding two features at 
a second epoch (epoch~B).  The difference between the two values
of $\Delta \theta$ is the pairwise separation and can be written
\begin{equation}
{\Delta \theta}_{\rm B} - {\Delta \theta}_{\rm A} = 
|{r_i - r_j}|_{\rm B} - |{r_i - r_j}|_{\rm A}\,, \qquad i=1,n; \quad j=i+1,n
\end{equation}
where $r_i = (x_i, y_i)$ and $(x_i, y_i)$ are the relative offsets in 
right ascension and declination, respectively.   The procedure is repeated 
for all possible pair combinations and the separations can be plotted as 
a histogram (e.g. Figure~\ref{PROP_HIST}).  The inclusion of all possible pair 
combinations often results in bimodal distribution with one of the peaks 
biased towards zero.  This is because some of the separations involve pairs of 
closely spaced maser components on the same side of the distribution 
(i.e. the northern or southern regions for OH\,12.8$-$0.9) that have 
little motion relative to one another.   For the 
sake of clarity, and to determine representative values for the angular shifts 
due to the expansion, we have included only those pairs separated by more 
than 80~mas in the histograms shown in Figure~\ref{PROP_HIST}.   
Figure \ref{PROP_HIST} plots the pairwise separations over time intervals of 
(a) 613, (b) 500 and (c) 114 days, respectively.  
All three histograms have centroids which are biased toward positive values
indicative of expansion.  The mean (median) differences between the 
north and south masers are 4.64 (4.69)~mas, 
3.88 (3.89)~mas, and 0.77 (0.80)~mas for 613, 500, and 114~days, respectively.  
The 1$\sigma$ standard deviations are 0.28, 0.21 and 0.13~mas, respectively.
The equivalent mean angular velocities are $2.76 \pm 0.16$~mas~yr$^{-1}$,
$2.83 \pm 0.15$~mas~yr$^{-1}$ and $2.47 \pm 0.41$~mas~yr$^{-1}$.  
To within the errors, the three velocities are consistent, and it is impossible
to determine whether there is any acceleration of the masers in the outflow. 

If 8~kpc is again assumed as the distance to OH\,12.8$-$0.9, the 
linear proper motions of the masers can be computed for the  
angular velocities above.  The resulting average linear
separation velocities are $105 \pm 6$~km~s$^{-1}$, 
$107 \pm 6$~km~s$^{-1}$ and $94 \pm 16$~km~s$^{-1}$  
for the 613, 500, and 114 day intervals respectively. 
If we assume that the north and south masers are moving outward 
at equal speeds (i.e. at half the computed separation velocity),
then the outflow velocity in the plane of the sky would be 
$\sim$53~km~s$^{-1}$ for the 613 day interval.  This velocity is 
slightly more than double the outflow velocity determined for 
OH\,12.8$-$0.9 in \cite{BM:05} based solely on the radial velocities.  
In the following section we update the radial velocities of the masers and 
combine them with the proper motions in order to estimate the 
full 3-D kinematics of the masers in the jet.

\subsection{3-D Maser Outflow \label{RAD_VEL}}

In \cite{BM:05}, we treated the radial velocity of OH\,12.8$-$0.9 
by simply determining the velocity separation of the blue- and red-shifted 
maser features representing the peaks in the spectral distribution.  This 
was an improvement over previous single-dish and VLA studies
that contained limited spatial information and likely suffered from blending of
features in the spectral domain.  In light of the two new epochs of VLBA
data, we felt that the more rigorous approach of computing spectral separations
($\Delta v$) from multiple features was more appropriate.  To do this 
we averaged the velocities of the blue- and red-shifted masers identified
in Tables~\ref{COMP_TABLE}--\ref{COMP_TABLE3} separately.  We
also computed the standard deviations for each set of components.  
From these averages we computed the spectral separation $\Delta v$
at each epoch.  The error in $\Delta v$ is simply the standard deviations added 
in quadrature.  Using this method, we find mean values for $\Delta v$ 
of $48.7 \pm 1.6$~km~s$^{-1}$,  $48.3 \pm 1.3$~km~s$^{-1}$ 
and $48.2 \pm 1.3$~km~s$^{-1}$ for epochs 1, 2, and 3, respectively.  
These values are also listed in Table~\ref{ACCEL_TABLE}. 

In \cite{BM:05} we compared the spectral separation of the 
blue- and red-shifted peaks in our spectrum with peak separations
from previous single dish and VLA observations
\citep[i.e.][]{ESW:86,GOMEZ:94, ENGELS:02}.   Here we wish
to compare our spectral separations computed from the component 
averages with these previous results, however, only \cite{ENGELS:02} 
tabulates parameters for individual spectral features.   \cite{ENGELS:02} 
categorizes components A--H of his Table 4 into the blue-shifted group
and components M--Q of Table 5 into the red-shifted group, with the 
remaining components, I--L, listed as intermediate.  From this 
information, we computed average velocities for the blue- and 
red-shifted spectral features and values for $\Delta v$ in a manner 
similar to that performed for our own VLBA data.  The resulting 
values of $\Delta v$ and the corresponding errors for the 
\cite{ENGELS:02} are listed in Table~\ref{ACCEL_TABLE}.  The 
only drawback to the Engels data as compared to our VLBA data
is the fact that there is no spatial information, thus the degree to 
which spectral blending is a factor is unknown.  It should be noted, 
that the spectral resolution of the Engels measurements is slightly 
better than our VLBA resolution at 0.16~km~s$^{-1}$.    

\begin{figure}[hbt]
\plotone{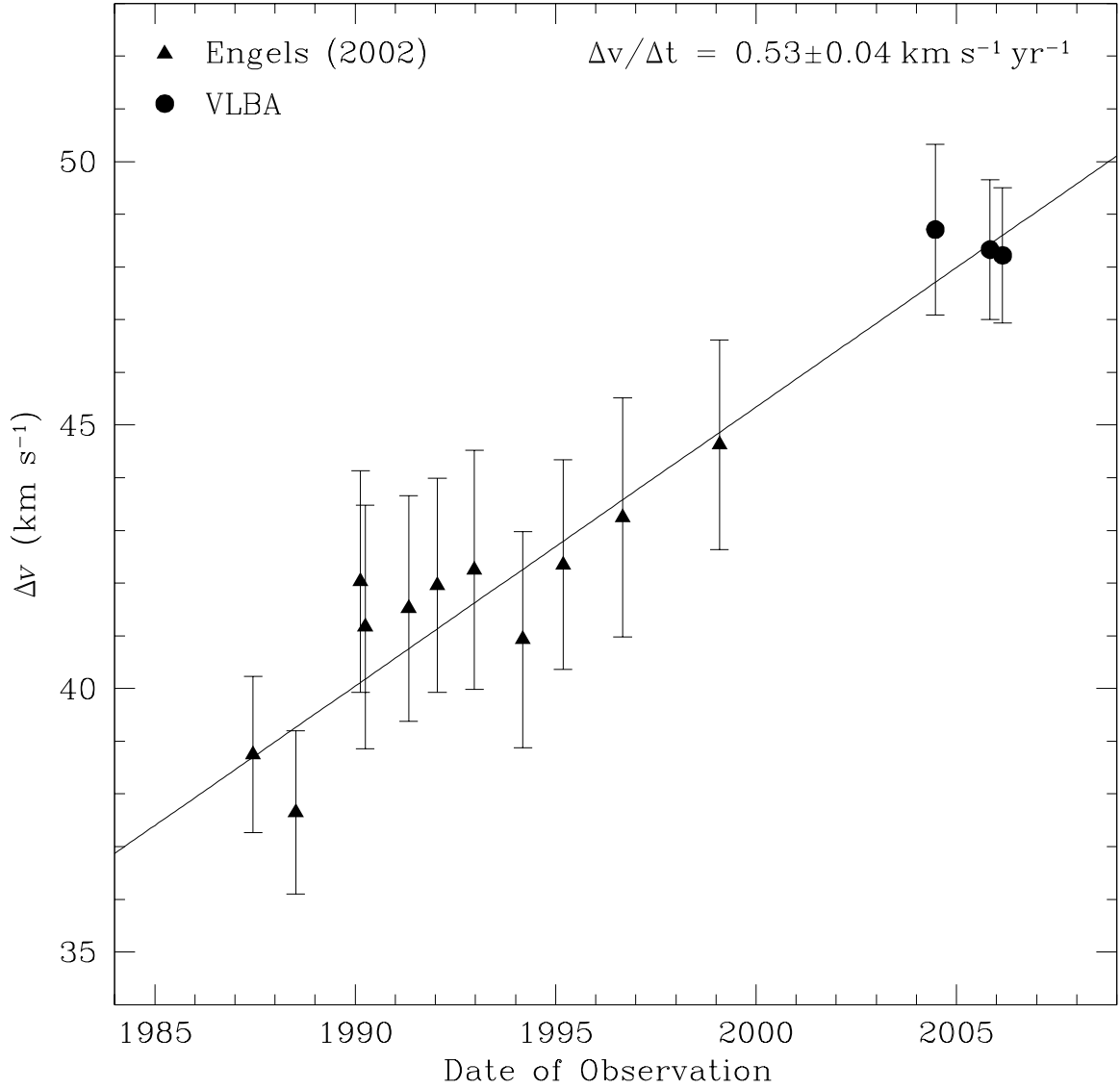}
\figcaption{Plot showing the measured velocity separation ($\Delta v$) 
of the blue- and red-shifted water maser features as a function of time from values 
listed in Table \ref{ACCEL_TABLE}.   The line represents a weighted
linear least-squares fit to the data and indicates a constant acceleration 
of 0.53\,km\,s$^{-1}$\,yr$^{-1}$. \label{ACCEL_PLOT}}
\end{figure}

Plotted in Figure~\ref{ACCEL_PLOT} are the results of our 
computation of $\Delta v$ from the blue- and red-shifted component 
averages.  The values from the 11 epochs of \cite{ENGELS:02} are 
plotted as filled triangles and those from our three epochs of VLBA data 
as filled circles.   The data clearly shows an increase in $\Delta v$ 
as a function of time.   To this data, we performed a linear least-squares
fit weighted by squares of the errors.  This fit is plotted as the line 
in Figure~\ref{ACCEL_PLOT}, with a slope of 
$0.53 \pm 0.04$~km~s$^{-1}$~yr$^{-1}$ corresponding to the 
relative acceleration of the blue- and red-shifted masers in 
the radial direction.  This is slightly lower than the 
$0.68 \pm 0.06$~km~s$^{-1}$~yr$^{-1}$ reported in \cite{BM:05}.   
With the new procedure of averaging over multiple spectral features, 
we have attempted to remove the error involved in computing the 
velocity separation from single peaks on the blue- and red-shifted 
sides of the spectrum that are likely unrelated over over periods
greater than the lifetime of the components $<3$\,yr \cite{ENGELS:02}.     
Engels also discusses $\sim$1~km~s$^{-1}$ velocity shifts that were 
observed for some components in the spectra of OH12.8$-$0.9 and the 
likelihood of these shifts being caused by intensity changes within 
blended features.   Such shifts could affect the velocity separations 
computed from single spectral peaks as was done in \cite{BM:05}.
 
If we assume an average over our three VLBA epochs, 
$\Delta v \approx 48$~km~s$^{-1}$, as the speed the masers are moving 
away from each other along the line of sight to the observer, then the outflow 
velocity in the radial direction is $\sim$24~km~s$^{-1}$.  
The 3-D outflow velocity of the masers may be estimated by combining the
radial motion above with the motion in the plane of the sky 
determined in \S~\ref{PROP_MOT}.  We find the 3-D outflow velocity of the 
masers to be $\sim$58~km~s$^{-1}$ with an inclination angle for the jet of 
$\sim$24$^{\circ}$ with respect to the plane of the sky.  The 3-D outflow 
acceleration, assuming this inclination angle and the radial acceleration 
above, is $\sim$0.63~km~s$^{-1}$~yr$^{-1}$.
The outflow acceleration in the plane of the sky would be 
$\sim$0.58~km~s$^{-1}$~yr$^{-1}$ under this assumption.  Thus 
over the 613 days between epochs 1 and 3, we could expect a change
in velocity of the blue-shifted masers relative to the red-shifted
masers of 1.93~km~s$^{-1}$ or 0.05~mas~yr$^{-1}$, which is 
undetectable in our present observations.

From the above information and two additional assumptions, namely: 
that the dynamical center of the outflow is the midpoint along the 
axis of separation between the two maser regions and that the 
masers have zero initial velocity, then an upper limit to the dynamical 
age of the outflow can be computed.   We find this dynamical age
to be $\sim$90~yr.  This value is roughly consistent with the age 
determined in \cite{BM:05}, which was estimated with no knowledge 
of the maser motions in the plane of the sky or the inclination angle 
of the outflow.  Assuming that the acceleration has and will remain 
constant at the above value of 0.63~km~s$^{-1}$~yr$^{-1}$, 
then the total time to reach an outflow velocity comparable to the other water 
fountain sources ($\Delta v \approx 150$~km~s$^{-1}$) is only about 
$\sim$240~yr, roughly 2.5 times the current dynamical age 
of OH\,12.8$-$0.9.  

In order to compare the characteristics of OH\,12.8$-$0.9 to the 
other water-fountain sources, we have summarized the relevant 
properties of all four sources in Table~\ref{CHAR_TABLE}. 
From the table we see that the assumed distance to OH\,12.8$-$0.9 is 
intermediate among the four sources.  The peak flux density is 
shown as a range measured over multiple epochs by \cite{ENGELS:02} 
for OH\,12.8$-$0.9 and by \cite{LMM:92} for the three other sources. 
The peak flux range for OH\,12.8$-$0.9 is similar to that 
of the other distant source IRAS\,19134+2131 and weaker than those 
of W\,43A and IRAS\,16342$-$3814.  The angular extent of the outflow
on the plane of the sky is also similar to that of
IRAS\,19134+2131.  The linear extent of OH\,12.8$-$0.9, however, is 
much less than that of any of the other water-fountain sources. 
An increase in the assumed distance to the source, would 
serve to bring this value in line with the other sources.  For the 
collimation of the OH\,12.8$-$0.9 jet, we have used an average of the 
northern and southern arcs and find that the collimation is slightly larger 
than for the other three water-fountain sources.  If we use only the southern
arc, or we disregard the one northern feature at (15.0, $-$10.0) then
the collimation is $\sim$10$^{\circ}$, which is consistent with the other 
water-fountain sources.   The estimated inclination angle of the outflow 
is also comparable values estimated for the other three objects.  

Aside from the difference in linear extent, the primary difference between
OH\,12.8$-$0.9 and the other water-fountain sources is in the kinematics.
Both the radial ($v_{rad}$) and tangential ($v_{tan}$) outflow velocities 
are less than half the values for the other sources.  It is possible that 
the tangential velocity could be closer to that of the other sources if the
distance to the source turns out to be greater than 8\,kpc.  An increased
distance would not increase the radial velocity estimate, but would instead
result in a decreased estimate for the inclination angle of the jet.  Another 
major difference between OH\,12.8$-$0.9 and the three other water 
fountains is that OH\,12.8$-$0.9  appears to be the only object with a measurable 
acceleration ($a_{rad}$) of the masers along the line of sight.
\cite{LMM:92} observed the H$_2$O maser spectra of the three other water-fountain
sources over the course of a few years and found no evidence for systematic
velocity drifts in spectral features that would suggest acceleration of the masers. 
The acceleration in combination with the smaller extent of the masers 
toward OH\,12.8$-$0.9 might suggest that it is younger than the other 
three objects.  This is not, however, confirmed by the computed dynamical
age of the jet, which is intermediate among the four sources at 90\,yr.

One characteristic that is not shown in the tables is the morphology
of the masers relative to the central star and the jet direction.  From the 
images of OH\,12.8$-$0.9 presented here, we find that the masers are 
arranged in point symmetric fashion along arcs perpendicular to the 
direction of the outflow, suggesting that the outflow is impacting a 
denser surrounding medium.
This bow shock structure is also observed for the H$_2$O masers
toward IRAS\,16342$-$3814 at least for the blue-shifted side of the jet
\citep{CSM:04}.  These two sources are in contrast to W\,43A where 
the masers are arranged along the direction of the outflow.  The structure 
of the W\,43A masers is well represented by a precessing jet model 
\citep{IMAI:02, IMAI:05}.  The remaining 
water-fountain source, IRAS\,19134+2131, shows the point symmetric
morphology of the three other sources, but the relationship between 
the masers and the outflow is less clear.  From the two epochs of 
VLBI observations reported in \cite{IMAI:04}, the masers on the eastern 
red-shifted side appear to be roughly aligned with the line 
connecting them to the western blue-shifted masers.  The motions 
of the masers, however, do not appear to be along this line, but 
rather perpendicular to this in a primarily north-south direction.  

Similar to the W\,43A outflow, precession is indicated for IRAS\,16342$-$3814,
but not as a result of the available H$_2$O maser data.   Instead, it was
the Keck adaptive optics images in the near-infrared \citep{SAHAI:05} that 
showed a corkscrew structure inscribed on the walls of the observed 
lobes.  The combined radio/infrared observations demonstrate that both 
bow shock morphologies and precession can take place in the same object.  
Since no such high-resolution optical/infrared imaging is available for 
OH\,12.8$-$0.9 it is unknown whether precession occurs in the object, 
however, it is clearly not suggested by the structure or kinematics of the 
H$_2$O masers.

\section{CONCLUSIONS}

Using the VLBA we have shown that, like the spatial morphology, the 
kinematics of the H$_2$O masers toward OH\,12.8$-$0.9 indicate
that the object is a member of the rare ``water fountain" class of sources.  
We find that the masers continue to be located in two distinct regions at the 
ends of a bipolar jet-like structure having a north-south orientation on 
the sky.  The masers in the two regions are distributed 
along arc-like structures $\sim$12--20\,mas across oriented roughly 
perpendicular to the separation axis.  The angular separation
of the two regions is roughly 110\,mas, with the blue-shifted masers located
to the north and the red-shifted masers to the south.   The arc-like arrangements 
of the masers are suggestive of bow shocks formed by a collimated 
axisymmetric wind impinging on the ambient medium surrounding the star.   

The two additional epochs of VLBA observations, beyond our first epoch
in 2004, have allowed us to track the projected spatial motions of individual features 
over the course of $\sim$600~days.  We find that the two maser regions 
maintain their arc-like appearance as they expand away from each other 
along the axis of separation.  The observed proper motion of the southern 
masers relative to the northern masers is 2.76~mas yr$^{-1}$ 
($\sim$105\,km\,s$^{-1}$ at the assumed distance of 8~kpc).  The masers
maintain collimation during this expansion with little motion perpendicular 
to the north--south axis.  Combining the proper motions with the 
measured radial velocities yields a three-dimensional (3-D) expansion 
velocity of  58\,km\,s$^{-1}$ relative to the dynamical center of the distribution.
The inclination angle of the jet is estimated to be $\sim$24$^{\circ}$. 
A simple linear fit to the historical radial velocity data combined with
our own measurements yields an estimate for the 3-D acceleration 
of $\sim$0.63~km~s$^{-1}$~yr$^{-1}$ and a corresponding dynamical 
age for the outflow of $\sim$90~yr.   Long-term VLBA monitoring of the 
H$_2$O masers should enable the determination of the 
tangential acceleration of the masers and the true 3-D acceleration of 
the outflow associated with OH\,12.8$-$0.9.  

\acknowledgements
This research has made use of the SIMBAD database, operated at CDS, 
Strasbourg, France.  This research has made use of NASA's Astrophysics Data 
System Bibliographic Services.

\clearpage

\begin{deluxetable}{ccccccc}
\tabletypesize{\footnotesize}
\tablewidth{0pt}
\tablecaption{H$_2$O maser characteristics derived from the epoch 1 (2004 June 21) 
VLBA data.
\label{COMP_TABLE}}
\tablehead{
\colhead{$v_{\rm LSR}$} & \colhead{S$_{\nu}$} & \colhead{$\sigma_{\rm S_{\nu}}$} & \colhead{Relative R.A.} &   
\colhead{$\sigma_{\rm R.A.}$} & \colhead{Relative Dec.} & \colhead{$\sigma_{\rm Dec.}$} \\
\colhead{(km\,s$^{-1}$)} & \colhead{(Jy beam$^{-1}$)} &  \colhead{(Jy beam$^{-1}$)} & \colhead{(mas)} &  
\colhead{(mas)}  & \colhead{(mas)} & \colhead{(mas)} \\
} 
\startdata
\multicolumn{7}{l}{Blue-shifted masers} \\
& & & & & & \\
$-$85.07  &  0.044  &  0.009  &  5.869  &  0.038  &  2.520  &  0.063 \\
$-$83.32  &   0.052 &  0.009  & 14.806 &  0.034  &  $-$10.133 & 0.049 \\
$-$81.71  &  1.297  &  0.012  &  0.000  &  0.006  &  0.000  &  0.008 \\
$-$81.67  &  0.123  &  0.012  &  $-$1.813  &  0.031  &  $-$0.322  &  0.045 \\
$-$80.88  &  0.261  &  0.009  &  7.647  &  0.009  &  0.376  &  0.014 \\
& & & & & & \\
\multicolumn{7}{l}{Red-shifted masers} \\
& & & & & & \\
$-$34.61  &  0.247  &  0.009  &  1.116  &  0.012  &  $-$108.821  &  0.019 \\
$-$33.92  &  0.139  &  0.012  &  $-$5.773  &  0.018  &  $-$108.460  &  0.027 \\
$-$33.72  &  0.498  &  0.012  &  2.235  &  0.018  &  $-$108.856  &  0.030 \\
$-$33.27  &  1.041  &  0.012  &  $-$2.807  &  0.005  &  $-$108.597  &  0.007 \\
$-$32.58  &  0.499  &  0.009  &  5.297  &  0.010  &  $-$106.906  &  0.017 \\
\enddata
\end{deluxetable}

\begin{deluxetable}{ccccccc}
\tabletypesize{\footnotesize}
\tablewidth{0pt}
\tablecaption{H$_2$O maser characteristics derived from the epoch 2 (2005 Nov. 2) VLBA data.
\label{COMP_TABLE2}}
\tablehead{
\colhead{$v_{\rm LSR}$} & \colhead{S$_{\nu}$} & \colhead{$\sigma_{\rm S_{\nu}}$} & \colhead{Relative R.A.} &   
\colhead{$\sigma_{\rm R.A.}$} & \colhead{Relative Dec.} & \colhead{$\sigma_{\rm Dec.}$} \\
\colhead{(km\,s$^{-1}$)} & \colhead{(Jy beam$^{-1}$)} &  \colhead{(Jy beam$^{-1}$)} & \colhead{(mas)} &  
\colhead{(mas)}  & \colhead{(mas)} & \colhead{(mas)} \\
} 
\startdata
\multicolumn{7}{l}{Blue-shifted masers} \\
& & & & & & \\
$-$83.29  &  0.153  &  0.009  &  15.458  &  0.029  &  $-$10.516  &  0.045 \\
$-$81.57  &  1.323  &  0.010  &  0.000  &  0.017  &  0.000  &  0.028 \\
$-$81.41  &  0.096  &  0.010  &  $-$2.207  &  0.041  &  $-$0.456  &  0.063 \\
$-$80.73  &  0.216  &  0.010  &  8.174  &  0.032  &  0.173  &  0.051 \\
& & & & & & \\
\multicolumn{7}{l}{Red-shifted masers} \\
& & & & & & \\
$-$34.47  &  0.468  &  0.010  &  1.286  &  0.020  &  $-$112.898  &  0.033 \\
$-$33.54  &  0.293  &  0.009  &  2.518  &  0.020  &  $-$112.793  &  0.034 \\
$-$33.31  &  0.086  &  0.009  &  $-$6.432  &  0.048  &  $-$112.695  &  0.071 \\
$-$33.10  &  0.569  &  0.009  &  $-$2.658  &  0.020  &  $-$112.669  &  0.030 \\
$-$32.68  &  0.142  &  0.009  &  5.765  &  0.034  &  $-$110.732  &  0.051 \\
\enddata
\end{deluxetable}

\begin{deluxetable}{ccccccc}
\tabletypesize{\footnotesize}
\tablewidth{0pt}
\tablecaption{H$_2$O maser characteristics derived from the epoch 3 (2006 Feb. 24) 
VLBA data.
\label{COMP_TABLE3}}
\tablehead{
\colhead{$v_{\rm LSR}$} & \colhead{S$_{\nu}$} & \colhead{$\sigma_{\rm S_{\nu}}$} & \colhead{Relative R.A.} &   
\colhead{$\sigma_{\rm R.A.}$} & \colhead{Relative Dec.} & \colhead{$\sigma_{\rm Dec.}$} \\
\colhead{(km\,s$^{-1}$)} & \colhead{(Jy beam$^{-1}$)} &  \colhead{(Jy beam$^{-1}$)} & \colhead{(mas)} &  
\colhead{(mas)}  & \colhead{(mas)} & \colhead{(mas)} \\
} 
\startdata
\multicolumn{7}{l}{Blue-shifted masers} \\
& & & & & & \\
$-$83.29  &  0.174  &  0.013  &  15.580  &  0.029  &  $-$10.741  &  0.054 \\
$-$81.54  &  0.970  &  0.013  &  0.000  &  0.026  &  0.000  &  0.043 \\
$-$81.30  &  0.079  &  0.013  &  $-$2.358  &  0.063  &  $-$0.384  &  0.110 \\
$-$81.26  &  0.065  &  0.013  &  0.347  &  0.068  &  1.034  &  0.111 \\
$-$80.72  &  0.094  &  0.012  &  8.208  &  0.043  &  0.067  &  0.074 \\
& & & & & & \\
\multicolumn{7}{l}{Red-shifted masers} \\
& & & & & & \\
$-$34.44  &  0.228  &  0.015  &  1.315  &  0.039  &  $-$113.770  &  0.072 \\
$-$33.46  &  0.151  &  0.014  &  2.582  &  0.039  &  $-$113.587  &  0.070 \\
$-$33.35  &  0.061  &  0.012  &  $-$6.378  &  0.062  &  $-$113.447  &  0.106 \\
$-$33.04  &  0.164  &  0.012  &  $-$2.593  &  0.041  &  $-$113.582  &  0.075 \\
$-$32.71  &  0.123  &  0.012  &  5.717  &  0.047  &  $-$111.546  &  0.085 \\
\enddata
\end{deluxetable}

\begin{deluxetable}{lcccc}
\tablewidth{0pt}
\tabletypesize{\footnotesize}
\tablecaption{
Velocity separation of blue- and red-shifted water masers.\label{ACCEL_TABLE}
}
\tablehead{
\colhead{Obs. Date} & \colhead{$\Delta\,v$} & \colhead{$\sigma_{\Delta\,v}$} &
\colhead{Spectral res.} & \colhead{} \\
\colhead{(yr)} & \colhead{(km\,s$^{-1}$)} & \colhead{(km\,s$^{-1}$)} & 
\colhead{(km\,s$^{-1}$)}  & \colhead{Ref.} 
} 
\startdata
1987.452 & 38.8 & 1.5 & 0.16 & 1 \\
1988.516 & 37.7 & 1.6 & 0.16 & 1 \\
1990.132 & 42.0 & 2.1 & 0.16 & 1 \\
1990.249 & 41.2 & 2.3 & 0.16 & 1 \\
1991.332 & 41.5 & 2.1 & 0.16 & 1 \\
1992.049 & 42.0 & 2.0 & 0.16 & 1 \\
1992.975 & 42.3 & 2.3 & 0.16 & 1 \\
1994.184 & 40.9 & 2.1 & 0.16 & 1 \\
1995.186 & 42.4 & 2.0 & 0.16 & 1 \\
1996.675 & 43.3 & 2.3 & 0.16 & 1 \\
1999.088 & 44.6 & 2.0 & 0.16 & 1 \\
2004.473 & 48.7 & 1.6 & 0.22 & 2 \\
2005.838 & 48.3 & 1.3 & 0.22 & 2 \\
2006.151 & 48.2 & 1.3 & 0.22 & 2 \\
\enddata
\tablecomments{References: (1) \cite{ENGELS:02}, (2) this work. }
\end{deluxetable}

\begin{deluxetable}{lcccc}
\tablewidth{0pt}
\tabletypesize{\footnotesize}
\tablecaption{
Summary of the H$_2$O maser characteristics for the water fountain 
sources.\label{CHAR_TABLE}
}
\tablehead{
\colhead{Characteristic} & \colhead{OH\,12.8$-$0.9} & 
\colhead{W\,43A} & \colhead{IRAS\,19134+2131} & 
\colhead{IRAS\,16342$-$3814} \\
} 
\startdata
Distance (kpc)  \dotfill                    & 8 & 2.6 & 16 & 2 \\
S$_{\rm peak}$ (Jy)  \tablenotemark{a} \dotfill & 1.4--13.0 & 11.5--56.6 &
1.2--6.0 & 16.0--67.7 \\
Angular Extent (mas) \tablenotemark{b} \dotfill & 110  & 920 &  135  & 3000 \\
Linear Extent (AU) \tablenotemark{b} \dotfill     &  880 & 2400 & 2180  & 6000 \\
Outflow Collimation (deg.) & 15 & 5 & 10 & 6 \\
Inclination Angle (deg.) \tablenotemark{c} \dotfill  & 24  & 39  & 25 &  40
\tablenotemark{d}  \\
Outflow Velocity:     &  &  &  &  \\                         
\phantom{or}$v_{\rm tan}$ (km\,s$^{-1}$) \dotfill & 53  & 110  & 120  & \nodata \\
\phantom{or}$v_{\rm rad}$ (km\,s$^{-1}$) \dotfill & 24 & 90 &  65  &  130 \\
\phantom{or}$v_{\rm tot}$  (km\,s$^{-1}$) \dotfill & 58 &  145 &  130 &  \nodata \\
Outflow Acceleration:     &  &  &  &  \\                         
\phantom{or}$a_{\rm rad}$ (km\,s$^{-1}$\,yr$^{-1}$) \dotfill &  0.53 &
\nodata   & \nodata  &  \nodata \\
Dynamical Age (yr)  \dotfill &  90  & 50 & 50  &  150 \tablenotemark{d} \\
References              \dotfill &  1, 2, 3  &  4, 5, 6  & 4, 7 & 4, 8, 9, 10 \\
\enddata
\tablenotetext{a}{Range over multiple epochs of single dish observations.}
\tablenotetext{b}{Measured in the plane of the sky.}
\tablenotetext{c}{Relative to the plane of the sky.}
\tablenotetext{d}{From OH maser observations.}
\tablecomments{References: (1) \cite{ENGELS:02}, (2) \cite{BM:05}, 
(3) this work, (4)~\cite{LMM:92}, (5)~\cite{IMAI:02}, (6)~\cite{IMAI:05}, 
(7)~\cite{IMAI:04}, (8)~\cite{STMZL:99}, (9)~\cite{MSC:03}, (10)~\cite{CSM:04}.}
\end{deluxetable}

\clearpage

\epsscale{0.9}
\epsscale{0.9}
\epsscale{1.0}
\epsscale{1.0}
\epsscale{1.0}
\epsscale{1.0}


\clearpage

\begin{thebibliography}{}

\bibitem[Aaquist \& Kwok(1996)]{AK:96}
Aaquist, O.~B. \& Kwok, S. 1996, \apj, 462, 813

\bibitem[Baud et al.(1979)]{BAUD:79}
Baud, B., Habing, H. J., Matthews, H. E. \& Winnberg, A. 
1979, \aaps, 36, 193

\bibitem[Baud et al.(1985)]{BAUD:85}
Baud, B., Sargent, A. J., Werner, M. W. \& Bentley, A. F. 
1985, \apj, 292, 628

\bibitem[Bloemhof et al.(1992)]{BRM:92} 
Bloemhof, E.E., Reid, M.J., \& Moran, J.M. 1992,
\apj, 397, 500

\bibitem[Boboltz \& Marvel(2005)]{BM:05}
Boboltz, D. A. and Marvel, K. B. 2005, \apjl, 627, 45

\bibitem[Boboltz et al.(1997)]{BDK:97} 
Boboltz, D.~A., Diamond, P.~J. \& Kemball, A.~J. 1997, 
\apjl, 487, L147

\bibitem[Chapman et al.(1991)]{CCS:91} 
Chapman, J.M., Cohen, R.J., \& Saika, D.J. 1991, 
\mnras, 249, 227

\bibitem[Chen et al.(2006)]{CSIK:06}
Chen, X., Shen, Z.-Q., Imai, H., \& Kamohara, R. 2006,
\apj, 640, 982 

\bibitem[Claussen et al.(2004)]{CSM:04} 
Claussen, M., Sahai, R., \& Morris, M.\ 2004, ASP Conf.~Ser.~313: 
Asymmetrical Planetary Nebulae III: Winds, Structure and the 
Thunderbird, 313, 331 

\bibitem[Condon(1997)]{CONDON:97}
Condon, J.J. 1997, \pasp, 109, 166

\bibitem[Engels(2002)]{ENGELS:02}
Engels, D. 2002, \aap, 388, 252

\bibitem[Engels et al.(1986)]{ESW:86}
Engels, D., Schmid-Burgk, J. \& Walmsley, C. M. 1986, \aap, 167, 129

\bibitem[G\'omez et al.(1994)]{GOMEZ:94}
G\'omez, Y., Rodriguez, L. F., Contr\'eras, M. E. \& Moran, J. M.
1994, Revista Mexicana de Astronomia y Astrofisica, 28, 97

\bibitem[Hodge et al.(2004)]{HKPW:04} 
Hodge, T.~M., Kraemer, K.~E., Price, S.~D., \& 
Walker, H.~J.\ 2004, \apjs, 151, 299 

\bibitem[Imai et al.(2004)]{IMAI:04}
Imai, H., Morris, M., Sahai, R. Hachisuka, K. \& Azzolini F, J. R.
2004, \aap, 420, 265
 
\bibitem[Imai et al.(2005)]{IMAI:05}
Imai, H., Nakashima, J.-i., Diamond, P.~J., Miyazaki, A., \& 
Deguchi, S.\ 2005, \apjl, 622, L125 

\bibitem[Imai et al.(2002)]{IMAI:02}
Imai, H., Obara, K., Diamond, P. J., Omodaka, T. \& Sasao, T. 
2002, Nature, 417, 829

\bibitem[Kemball(1992)]{KEMBALL:92} 
Kemball, A.J. 1992, Ph.D. Thesis, Rhodes Univ. 

\bibitem[Kwok(1993)]{KWOK:93} 
Kwok, S.\ 1993, \araa, 31, 63

\bibitem[Kwok et al.(1997)]{KVB:97}
Kwok, S., Volk, K. \& Bidelman, W.~P.\ 1997, \apjs, 112, 557

\bibitem[Likkel et al.(1992)]{LMM:92}
Likkel, L., Morris, M. \& Maddalena, R.~J.\ 1992, \aap, 256, 581 

\bibitem[Manchado et al.(2000)]{MVSG:00} 
Manchado, A., Villaver, E., Stanghellini, L., \& Guerrero, M.~A.\ 2000, 
ASP Conf.~Ser.~199: Asymmetrical Planetary Nebulae II: From Origins to 
Microstructures, 199, 17 

\bibitem[Marvel(1996)]{MARVEL:96} 
Marvel, K.B.\ 1996, Ph.D. Thesis, New Mexico State Univ.

\bibitem[Morris et al.(2003)]{MSC:03}
Morris, M.~R., Sahai, R. \& Claussen, M.\ 2003, RevMexAA, 
15, 20

\bibitem[Sahai et al.(1999)]{STMZL:99}
Sahai, R., Te Lintel Hekkert, P., Morris, M., Zijlstra, A., \& Likkel, L.\ 1999, 
\apjl, 514, L115 

\bibitem[Sahai \& Trauger(1998)]{ST:98}
Sahai, R. \& Trauger, J.~T.\ 1998, \aj, 116, 1357

\bibitem[Sahai et al.(2005)]{SAHAI:05}
Sahai, R., Le Mignant, D., S{\'a}nchez Contreras, C., 
Campbell, R.~D., \& Chaffee, F.~H.\ 2005, \apjl, 622, L53 

\bibitem[te Lintel Hekkert et al.(1989)]{teLINTEL:89}
te Lintel Hekkert, P., Versteege-Hensel, H.~A., Habing, H.~J. \& Wiertz, M.\
1989, \aaps, 78, 399 

\end{thebibliography}
\end{document}